\documentclass[%
 aip,
 amsmath,amssymb,
 reprint,%
]{revtex4-1}

\renewcommand{\d}{\boldsymbol{d}}

\usepackage{braket}
\usepackage{graphicx}
\usepackage{dcolumn}
\usepackage{bm}

\usepackage[utf8]{inputenc}
\usepackage[T1]{fontenc}
\usepackage{mathptmx}
\usepackage{amsmath}
\usepackage{etoolbox}
\usepackage{xcolor}
\usepackage{ulem}
\usepackage{siunitx}
\makeatletter
\def\@email#1#2{%
 \endgroup
 \patchcmd{\titleblock@produce}
  {\frontmatter@RRAPformat}
  {\frontmatter@RRAPformat{\produce@RRAP{*#1\href{mailto:#2}{#2}}}\frontmatter@RRAPformat}
  {}{}
}%
\makeatother
\begin{document}

\preprint{AIP/123-QED}

\title{Floquet Nonadiabatic Mixed Quantum-Classical Dynamics in Laser-Dressed Solid Systems}
\author{Jingqi Chen}
\affiliation{Fudan University, 220 Handan Road, Shanghai 200433, China}
 \affiliation{Department of Chemistry, School of Science, Westlake University, Hangzhou 310024 Zhejiang, China}
 \affiliation{Institute of Natural Sciences, Westlake Institute for Advanced Study, Hangzhou 310024 Zhejiang, China}
\author{Yu Wang}
\email{wangyu19@westlake.edu.cn}
 \affiliation{Department of Chemistry, School of Science, Westlake University, Hangzhou 310024 Zhejiang, China}
 \affiliation{Institute of Natural Sciences, Westlake Institute for Advanced Study, Hangzhou 310024 Zhejiang, China}
\author{Wenjie Dou}%
 \email{douwenjie@westlake.edu.cn}
 \affiliation{Department of Chemistry, School of Science, Westlake University, Hangzhou 310024 Zhejiang, China}
 \affiliation{Institute of Natural Sciences, Westlake Institute for Advanced Study, Hangzhou 310024 Zhejiang, China}
 \affiliation{Department of Physics, School of Science, Westlake University, Hangzhou 310024 Zhejiang, China}

\date{\today}

\begin{abstract}
In this paper, we introduce the Floquet Ehrenfest and Floquet surface hopping approaches to study the nonadiabatic dynamics in the laser-dressed solid systems. We demonstrate that these two approaches can be formulated in both real and reciprocal spaces. Using the two approaches, we are able to simulate the interaction between electronic carriers and phonons under periodic drivings, such as strong light-matter interactions. Employing the Holstein and Peierls models, we show that the strong light-matter interactions can effectively modulate the dynamics of electronic population and mobility. Notably, our study demonstrates the feasibility and effectiveness of modeling low-momentum carriers' interactions with phonons using a truncated reciprocal space basis, an approach impractical in real space frameworks. Moreover, we reveal that even with significant truncation, carrier populations derived from surface hopping maintain greater accuracy compared to those obtained via meanfield dynamics. These results underscore the potential of our proposed methods in advancing the understanding of carrier-phonon interactions in various laser-dressed materials.
\end{abstract}

\maketitle

\section{\label{sec:level1}Introduction}
In solid-state materials, various interactions such as electron-electron, electron-phonon, electron-hole, electron-impurity, and electron-photon interactions play pivotal roles in determining material properties and designing functional materials for diverse applications. Among these interactions, electron-phonon and electron-photon interactions are particularly significant for understanding carrier dynamics in extended systems.

Electron-phonon interactions significantly influence the electronic and thermal characteristics of solid-state materials.
For example, in hybrid metal halide perovskites, interactions between charge carriers and optical phonons dictate the broadening of emission lines at room temperature\cite{wright2016electron}.
Intense electron-phonon interactions in blue-emitting perovskites result in rapid non-radiative decay, thereby reducing the photoluminescence quantum yield\cite{gong2018electron}.
Furthermore, electron-phonon coupling is also a key mechanism for conventional superconductivity, where Cooper pairs of electrons are formed due to attractive interactions mediated by lattice vibrations (phonons)\cite{kulic2000interplay,shen2002role}. Experimental evidence suggests that electron-phonon coupling strongly influences the electron dynamics in high-temperature superconductors\cite{lanzara2001evidence,reznik2006electron}. 

Nowadays, strong light-matter interactions have become prominent, leading to significant and observable effects. 
With the development of modern techniques, properties of solids have been modified by light-matter strong couplings or laser fields \cite{garcia2021manipulating}.
For example, in the field of quantum computation, researchers have proposed a scheme to accelerate the nontrivial two-qubit phase gate by ultrastrongly coupling superconducting flux qubits to a circuit quantum electrodynamics system\cite{wang2017ultrafast, schiro2021quantum, cabra2020optical, tiwari2023floquet}. 
Experiments have demonstrated that materials subjected to intense terahertz pulses exhibit transient superconducting properties at significantly higher temperatures than at equilibrium \cite{mankowsky2014nonlinear,yang2019lightwave}.
Applying electromagnetic radiation to a topologically trivial insulator, such as a non-inverted HgTe/CdTe quantum well with no edge state in the static limit, can induce novel topological edge states \cite{lindner2011floquet}.
Suitably chosen irradiation parameters can open gaps at the Dirac point of graphene and even drive graphene into the topological Haldane phase \cite{kitagawa2010topological,kitagawa2011transport}.

From theoretical perspective, the analysis of periodically driven systems often relies on Floquet theorem \cite{bukov2015universal}.
Floquet formalism allows for the reduction of the periodic or quasi-periodical time-dependent Hamiltonian into time-independent Floquet matrix \cite{chu2004beyond, mori2023floquet, sato2020floquet}.
Several theoretical methods have been developed under the Floquet representation to solve problems related to periodically driven solid states.
For example, Kitagawa $et$ $al$. \cite{kitagawa2011transport} demonstrated that quantum transport properties can be controlled in materials such as graphene and topological insulators by applying light using Floquet Green's function. 
Tsuji $et$ $al$. \cite{tsuji2008correlated} studied correlated electron systems that are periodically driven by external fields via Floquet dynamical mean-field theory and showed that photoinduced midgap states emerge from strong ac fields, triggering an insulator-metal
transition.

In this study, we develop two trajectory-based nonadiabatic mix quantum-classical dynamics methods, namely the Floquet meanfield dynamics method (FMD) and the Floquet surface hopping (FSH) algorithm, for solid states under external periodic drivings.
Our research builds upon the foundation laid by existing real and reciprocal space semi-classical meanfield dynamics and surface hopping methods, as documented in Refs. [\!\!\citenum{krotz2021reciprocal}, \citenum{krotz2022reciprocal}]. 
Furthermore, our approach to Floquet mixed quantum-classical dynamics is directly connected to the Floquet quantum-classical Liouville equation, which has been rigorously derived in Ref. [\!\!\citenum{mosallanejad2023floquet}]. Note also that the connection between quantum-classical Liouville equation and mix quantum-classical dynamics, such as surface hopping has been established in Ref. [\!\!\citenum{subotnik2013can}]. 
While Floquet surface hopping has been employed to investigate mixed quantum-classical dynamics under periodic driving in the gas phase, see Ref. [\!\!\citenum{fiedlschuster2016floquet}]; our current study focuses on implementing FSH and FMD for solids.

We consider the interactions between electronic carriers and phonons both in real and reciprocal spaces.
Real space simulations are generally more suitable for disordered solids, while reciprocal space simulations are more suitable for ordered solids. 
Using a tight-binding model involving Holstein-type \cite{holstein1959studies,freericks1993holstein,zhang1999dynamical} and Peierls-type \cite{su1979solitons} electron-phonon couplings, we calculate electronic dynamics and charge mobility under different drivings.
We expect that the FMD and FSH algorithms, both in real and reciprocal space, can provide theoretical predictions on solid state property modified by external periodic drivings.  

The structure of this paper is organized as follows:
In Sec. \uppercase\expandafter{\romannumeral2},
we present the theory of general FMD and FSH in real- and reciprocal-space.
In Sec. \uppercase\expandafter{\romannumeral3},
we employ Holstein and Peierls models to simulate electronic dynamics and charge mobility under different periodic drivings using FMD and FSH. We consider the truncation in the Brillouin zone as well.
Finally, we conclude in Sec. \uppercase\expandafter{\romannumeral4}.

\section{Theory}
\subsection{Model Hamiltonian}
Without loss of generality, we focus on the one-dimensional lattice models that feature a solitary quantum state at each site within a unit cell. The total Hamiltonian for the coupled electron-phonon system gives as:
\begin{equation}\label{H_total}
    \hat{H}_{\text{tot}}(t) = \hat{H}_{\text{ph}}(\mathbf{R}, \mathbf{P}) + \hat{H}_{\text{el}}(t) + \hat{H}_{\text{el-ph}}(\mathbf{R}).
\end{equation}
In this formulation, the term $\hat{H}_{\text{ph}}(\mathbf{R}, \mathbf{P})$ denotes the classical Hamiltonian for the phonon energy, with $\mathbf{R} = (r_1, r_2, ..., r_N)$ ($N$ represents the total number of sites) and $\mathbf{P} = (p_1, p_2, ..., p_N)$ being the classical position and momentum vectors, respectively. The operator $\hat{H}_{\text{el}}(t)=\hat{H}_{\text{el}}(t+T)$ is the electronic Hamiltonian subjected to periodic drivings with a time period $T$. In addition, $\hat{H}_{\text{el-ph}}(\mathbf{R})$ represents the electron-phonon coupling. 

For a one-dimensional lattice of harmonic, noninteracting, and dispersionless phonons \cite{rosati2015dispersionless, glebov2000helium}, the real space Hamiltonian is given by:
\begin{equation}
H_{\text{ph}} = \sum_{n=1}^{N} \left(\frac{1}{2} p_n^2 + \frac{1}{2} \omega^2 r_n^2\right),
\end{equation}
where $\omega$ indicates the phonon energy. 
We remove the hat for $H_{\text{ph}}$ since we make classical approximation here. 
Note that our methods developed here are not restricted to one phonon frequency. Our methods can be used to study multiple phonons with different frequencies easily.
The transformation from real to reciprocal space gives as:
\begin{equation}
    \tilde{f}_k = \frac{1}{\sqrt{N}}\sum_ne^{ikn}f_n,
\end{equation}
Here, we set the lattice constant to be unit, and $k$ is the wavevector ranging from $-\pi$ to $\pi$. Henceforth, a tilde will be used to denote reciprocal-space coordinates.
After introducing classical equivalent of the ladder operator:
\begin{equation}
    \tilde{a}_k = \sqrt{\frac{\omega}{2}}(\tilde{r}_k+i\frac{\tilde{p}_k}{\omega}),
\end{equation}
we can get the position $\tilde{r}_k$ and momentum $\tilde{p}_k$ in the reciprocal space as \cite{krotz2021reciprocal, krotz2022reciprocal}:
\begin{equation}\label{rk}
\begin{split}
    \tilde{r}_k &= \frac{1}{\sqrt{2\omega}}(\tilde{a}_k+\tilde{a}_k^*) \\ &
    =\frac{1}{\sqrt{N}}\sum_{n=1}^{N} \left(r_n \cos(kn)-\frac{p_n}{\omega}\sin(kn)\right),
\end{split}
\end{equation}
\begin{equation}\label{pk}
\begin{split}
    \tilde{p}_k &= -i\sqrt{\frac{\omega}{2}}(\tilde{a}_k-\tilde{a}_k^*) \\ &
    =\frac{\omega}{\sqrt{N}}\sum_{n=1}^N \left(\frac{p_n}{\omega}\cos(kn)+r_n\sin(kn)\right).
\end{split}
\end{equation}
Note that under the transformation from real to reciprocal space, classical position and momentum coordinates become scrambled. Then, we obtain the phonon Hamiltonian in the reciprocal space with classical approximation:
\begin{equation}
    H_{\text{ph}} = \sum_{k = -\pi}^{\pi} \left(\frac{1}{2} \tilde{p}_k^2 + \frac{1}{2} \omega^2 \tilde{r}_k^2\right).
\end{equation}
For the electronic part, we consider a simple tight-binding model with periodic driving in real and reciprocal space:
\begin{equation}
\begin{split}
    \hat{H}_{\text{el}}(t) = & -(J_0 + J_1 \cos (\Omega t)) \sum_n  \left(\hat{c}_{n+1}^{\dagger} \hat{c}_n +\mathrm{h.c.}\right) \\
    = & -2(J_0 + J_1\cos (\Omega t)) \sum_k \hat{c}_{k}^{\dagger} \hat{c}_k \cos(k),
\end{split}
\end{equation}
where $\hat{c}_n^{(\dagger)}$ and $\hat{c}_k^{(\dagger)}$ are real and reciprocal space electronic annihilation (creation) operators, respectively. $(J_0 + J_1 \cos (\Omega t))$ is the nearest-interaction term. Note the light-matter interactions is introduced where the light couples to transition dipole between two sites: 
$J_0$ signifies the coupling strength inherent to the electronic states, while $J_1 \cos (\Omega t)$ provides a semi-classical portrayal of light-matter interactions: $J_1$ is the coupling strength, and $\Omega$ is the driving frequency. This coupling has been investigated in Refs.  [\!\!\citenum{asboth2014chiral,dal2015floquet}], which can be achieved in experiments, for example, by periodically bending an ultrathin metallic array of coupled corrugated waveguides \cite{cheng2019observation}. 

Then we give the Hamiltonian of electron-phonon coupling $\hat{H}_{\text{el-ph}}$ for two models. Firstly, as for the Holstein model \cite{krotz2021reciprocal, krotz2022reciprocal}:
\begin{eqnarray}
\begin{aligned}
    \hat{H}_{\text{el-ph}} =& g \sqrt{2 \omega^3} \sum_n \hat{c}_n^{\dagger} \hat{c}_n r_n \\
    = & g \sqrt{\frac{\omega}{2N}}\sum_{k, \kappa} \hat{c}_{k+\kappa}^{\dagger} \hat{c}_k (\omega (\tilde{r}_{-\kappa} + \tilde{r}_{\kappa}) - i (\tilde{p}_{-\kappa} - \tilde{p}_{\kappa})),
\end{aligned}
\end{eqnarray}
while for the Peierls model:
\begin{eqnarray}
\begin{aligned}
\hat{H}_{\text{el-ph}} =& g \sqrt{2\omega^3} \sum_n (\hat{c}_{n}^{\dagger} \hat{c}_{n+1} + \hat{c}_{n+1}^{\dagger} \hat{c}_{n}) (r_n -r_{n+1} ) \\
=& g \sqrt{\frac{2\omega}{N}}\sum_{k, \kappa} \hat{c}_{k+\kappa}^{\dagger} \hat{c}_k (i\omega (\tilde{r}_{-\kappa} + \tilde{r}_{\kappa}) + (\tilde{p}_{-\kappa} - \tilde{p}_{\kappa})) \\ & \times (\sin(k+\mathbf{\kappa})-\sin(k)).
\end{aligned}
\end{eqnarray}
Here, $g$ is the dimensionless coupling parameter, which is related to the vibrational reorganization energy as $g^2\omega$.

\subsection{Floquet Theory}
Floquet theory is a useful tool to address the complexities arising from the time-periodic Hamiltonian $\hat{H}(\mathbf{R},t) = \hat{H}(\mathbf{R},t+T)$. 
Consider a general periodic Hamiltonian and the time-dependent Schr{\"o}dinger equation:
\begin{eqnarray}
\label{eq1}
\begin{aligned}
    i\hbar \ket{\dot{\Psi}(\mathbf{R},t)} = \hat{H}(\mathbf{R},t) \ket{\Psi(\mathbf{R},t)}.
\end{aligned}    
\end{eqnarray}
Then we can expand our wavefunction in the following form as one solution to Eq. (\ref{eq1}), which is commonly used to describe energy eigenstates and their evolution:
\begin{equation}
    \Psi(\mathbf{R},t) = u(\mathbf{R},t)e^{-i\epsilon t / \hbar}.
\end{equation}
Here, $u(\mathbf{R},t)$ is a complex function that represents the time-varying amplitude of the wave function.
$e^{-i\epsilon t/\hbar}$ is a complex exponential function that represents the time-varying phase of the wave function,
$\epsilon$ is a constant representing an energy eigenvalue. If we take the expanded form of this wave function into Eq. (\ref{eq1}), we can get:
\begin{equation}
    H^{\text{F}} u(\mathbf{R},t) = \epsilon u(\mathbf{R},t),
\end{equation}
where $H^{\text{F}}(\mathbf{R},t) = \hat{H}(\mathbf{R},t) - i\hbar \frac{\partial}{\partial t}$ is called Floquet Hamiltonian. 
We aim to make Hamiltonian time-independent; the first step is to use the periodicity: $u(\mathbf{R},t) =  u(\mathbf{R},t+T)$, so it can be Fourier transformed as:
\begin{equation}
    u(\mathbf{R},t) = \sum_{\alpha = -n_{\text{F}}}^{n_{\text{F}}} c_{\alpha}(\mathbf{R}) e^{i\alpha \Omega t},
\end{equation}
where the frequency is $\Omega = \frac{2\pi}{T}$. In the next step, we do the unitary transform on the time-dependent Hamiltonian, and we have:
\begin{eqnarray}
\begin{aligned}
    &u^{\dagger}(\mathbf{R}, t) H^{\text{F}}(\mathbf{R},t) u(\mathbf{R}, t)   = \sum_{\beta=-m_{\text{F}}}^{m_{\text{F}}} \sum_{\alpha=- n_{\text{F}}}^{n_{\text{F}}} \\ &(c^{\dagger}_{\beta}(\mathbf{R}) e^{-i \beta \Omega t} \hat{H}(\mathbf{R},t) e^{i \alpha \Omega t} c_{\alpha}(\mathbf{R})
    - c^{\dagger}_{\beta}(\mathbf{R}) c_{\alpha}(\mathbf{R}) \alpha \hbar \Omega e^{i(\alpha-\beta)\Omega t}).
\end{aligned}
\end{eqnarray}
We need to do the integration in one time period to get the results:
\begin{equation}
\label{eq_time_average}
\begin{split}
    [ (\sum_{\beta=-m_{\text{F}}}^{m_{\text{F}}} \sum_{\alpha=-n_{\text{F}}}^{n_{\text{F}}} &\frac{1}{T} \int_{0}^{T}dt e^{-i\beta \Omega t} \hat{H}(\mathbf{R},t) e^{i\alpha \Omega t})\\ & - \alpha \hbar \Omega \delta_{\alpha \beta} ]c_{\alpha}(\mathbf{R}) = \epsilon c_{\beta}(\mathbf{R}).
\end{split}
\end{equation}
The procedure described in Eq. (16) corresponds to representing the equation of motion in Floquet space, where we can define the Floquet Hamiltonian: 
\begin{equation}
    \hat{H}^{\text{F}}(\mathbf{R}) = \sum_{\alpha} H_{\beta \alpha}(\mathbf{R}) - \alpha \hbar \Omega \delta_{\alpha \beta}.
\end{equation}
Note that the Hamiltonian in Floquet representation is time-independent. The time-dependency is incorporated in the replicas in the Floquet space. Note also that one can obtain the averaged observable directly in the Floquet representation (e.g., see Eqs. (\ref{eq21},\ref{eq22})).

Now, we can transform our model Hamiltonian into the Floquet representation. 
The Floquet theory used in this paper is a classical theory of light, which has the connection with optical cavity and can be used to simulate quantum electrodynamics \cite{hubener2021engineering}.
As for the Holstein model, we can construct each single block $H_{\beta \alpha }$ as:
\begin{equation}
\begin{aligned}
    H_{ \beta \alpha} =&  [-J_0 \sum_n \left(\hat{c}_{n+1}^{\dagger} \hat{c}_n + \text{h.c.}\right) + g\sqrt{2\omega^3} \sum_n \hat{c}_n^{\dagger}\hat{c}_n r_n \delta_{\alpha \beta}  \\
    & - \frac{1}{2} J_1 \sum_n \left(\hat{c}_{n+1}^{\dagger} \hat{c}_n + \text{h.c.}\right) (\delta_{\alpha,\beta+1} + \delta_{\alpha,\beta-1})    \\
 =& [-2 J_0 \sum_k \hat{c}_{k}^{\dagger} \hat{c}_k \cos(k)+ g \sqrt{\frac{\omega}{2N}}\sum_{k, \kappa} \hat{c}_{k+\kappa}^{\dagger} \hat{c}_k (\omega (\tilde{r}_{-\kappa} + \tilde{r}_{\kappa}) \\
    &- i (\tilde{p}_{-\kappa} - \tilde{p}_{\kappa}))] \delta_{\alpha \beta} - J_1 \sum_k \hat{c}_{k}^{\dagger} \hat{c}_k \cos(k) (\delta_{\alpha,\beta+1} + \delta_{\alpha,\beta-1}),
\end{aligned}
\end{equation}
while, as for the Peierls model:
\begin{equation}
\begin{aligned}
    H_{ \beta \alpha}  =&  [ -J_0 \sum_n \left(\hat{c}_{n+1}^{\dagger} \hat{c}_n + \mathrm{h.c.}\right) \\ 
    &+ g \sqrt{2\omega^3} \sum_n (\hat{c}_{n}^{\dagger} \hat{c}_{n+1} + \hat{c}_{n+1}^{\dagger} \hat{c}_{n}) (r_n -r_{n+1} )] \delta_{\alpha \beta}  \\
    & - \frac{1}{2} J_1 \sum_n \left(\hat{c}_{n+1}^{\dagger} \hat{c}_n + \text{h.c.}\right) (\delta_{\alpha,\beta+1} + \delta_{\alpha,\beta-1}) \\
    =& [-2 J_0 \sum_k \hat{c}_{k}^{\dagger} \hat{c}_k \cos(k)+g \sqrt{\frac{2\omega}{N}}\sum_{k, \kappa} \hat{c}_{k+\kappa}^{\dagger} \hat{c}_k (i\omega (\tilde{r}_{-\kappa} + \tilde{r}_{\kappa}) \\
    &+ (\tilde{p}_{-\kappa} - \tilde{p}_{\kappa})) (\sin(k+\mathbf{\kappa})-\sin(k)) ] \delta_{\alpha \beta}\\ &- J_1 \sum_k \hat{c}_{k}^{\dagger} \hat{c}_k \cos(k) (\delta_{\alpha,\beta+1} + \delta_{\alpha,\beta-1}).
\end{aligned}
\end{equation}

\begin{figure*}[htbp]
\centering
\includegraphics[scale=0.5]{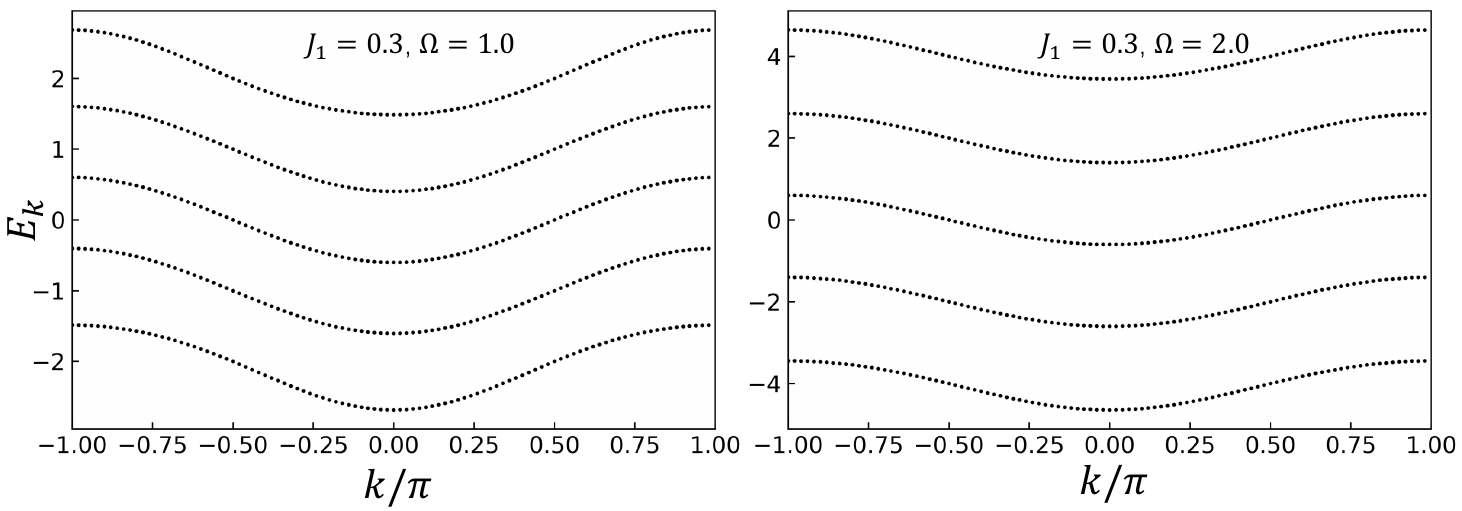}
\caption{Floquet replicas without electron-phonon coupling in reciprocal space. Here, we set the Floquet levels $n_{\text{F}} = 2$, so there are $5$ quasi-surfaces in total. Other parameters: $J = 0.3$, $\hbar\omega = 0.3$, $g = 0.5$, and $kT = 0.5$. Here, the energy unit corresponds to $50.5$ $meV$.}
\label{fig:floquet_replicas}
\end{figure*}

It is easier to understand the band structure in the reciprocal space according to its mathematical form. Therefore, we plot the multiple Floquet quasi-surfaces with different driving conditions in the first Brillouin zone, which are shown in Fig. \ref{fig:floquet_replicas}. 
The original band has been extended into multiple Floquet bands due to the Floquet drivings. The energy gap between the nearest two Floquet bands is proportional to the driving frequency. 


\subsection{Floquet Meanfield Dynamics (FMD)}
In this subsection, we will introduce the Floquet meanfield approach in real and reciprocal space.
In Ref. [\!\!\citenum{ivanov2021floquet}], it has been demonstrated that the equation of motion of the density operator obeys Liouville-von Neumann equation in Floquet representation. Therefore, we can deduce the following equation:
\begin{equation}
    \frac{\partial}{\partial t}\ket{\Psi^{\text{F}}(t)} = - i \hbar \hat{H}^{\text{F}}(\mathbf{R}) \ket{\Psi^{\text{F}}(t)}.
\end{equation}
Here, $\hat{H}^{\text{F}}(\mathbf{R})$ including pure electron Hamiltonian and electron-phonon coupling Hamiltonian, while $\Psi^{\text{F}}$ represents the Floquet wavefunction. In Floquet space, each original surface has multiple Floquet replicas with energy gap of $\alpha \hbar \Omega$, where $\alpha$ is an integer ranging from negative infinity to positive infinity. Initially, we sample all trajectories on $\alpha=0$ Floquet state. We can express the Ehrenfest dynamics in real space as follows \cite{krotz2021reciprocal}:
\begin{eqnarray}
\label{eq21}
\begin{aligned}
    &\dot r_n =  p_n, \\
    &\dot p_n =  - \omega^2 r_n - \langle \Psi^{\text{F}} \mid \nabla_{r_n} \hat{H}^{\text{F}} \mid \Psi^{\text{F}} \rangle.
\end{aligned}
\end{eqnarray}
Next, we will extend this approach to the dynamical evolution in reciprocal space as well.
The evolution of wavefunctions and nuclear coordinates in reciprocal space follows a similar approach as in real space. However, the classical position coordinates in real space depend both on position and momentum coordinates in reciprocal space as shown in Eqs. \ref{rk} and \ref{pk}.
Therefore, the equations of motion involve both position and momentum-derivative terms in reciprocal space are given by:
\begin{eqnarray}
\label{eq22}
\begin{aligned}
    &\dot r_k  =  p_k  +  \langle \Psi^{\text{F}} \mid \nabla_{p_k} \hat{H}^{\text{F}} \mid \Psi^{\text{F}} \rangle, \\
    &\dot p_k = - \omega^2 r_k - \langle \Psi^{\text{F}} \mid \nabla_{r_k} \hat{H}^{\text{F}} \mid \Psi^{\text{F}} \rangle.
\end{aligned}
\end{eqnarray}

\subsection{Floquet Surface Hopping(FSH)}
Within the surface hopping algorithm, one propagates a trajectory along the adiabatic surface, and integrates the electronic density matrix. Our FSH method is performed among multiple Floquet electronic orbitals, the hopping processes are elaborated in Ref. [\!\!\citenum{krotz2022reciprocal}].Here, we propagate a trajectory along the Floquet adiabatic surface. 
The hopping rate from surface $J$ to surface $I$ can be determined by\cite{subotnik2016understanding,wang2016recent, krotz2022reciprocal}:
\begin{equation}
    k_{J\rightarrow I} = \Theta(\Gamma_{J\rightarrow I}),
\end{equation}
where $\Theta$ is the Heaviside function:
\begin{equation}
    \Theta(x) = 
    \begin{cases}
        x & \text{x>0} \\
        0 & \text{x<0},
    \end{cases}
\end{equation}
and
\begin{equation}\label{Gamma_real}
    \Gamma_{J\rightarrow I} = -2\Re\left(\mathbf{P}\cdot\d_{IJ}\frac{\hat{\rho}^{\text{F(ad)}}_{JI}}{\hat{\rho}^{\text{F(ad)}}_{JJ}}\right).
\end{equation}
Here, we set nuclear mass and $\hbar$ equal to $1$. $\hat{\rho}^{\text{F(ad)}}=\ket{\Psi^{\text{F(ad)}}}\bra{\Psi^{\text{F(ad)}}}$ is the density matrix in adiabatic representation. $\d_{IJ}$ is the derivative coupling:
\begin{equation}
    \d_{IJ} = \frac{-\bra{\Psi_I^{\text{F(ad)}}}\frac{\partial \hat{H}^{\text{F}}}{\partial r_n}\ket{\Psi_J^{\text{F(ad)}}}}{E_I-E_J},
\end{equation}
where $\ket{\Psi^{\text{ad}}}$ and $E$ are the eigenvector and eigenvalue of $\hat{H}^{\text{F}}$, respectively.

After a hop from Floquet state $J$ to Floquet state $I$, one should rescale the nuclear momentum in the direction of derivative coupling $\d_{JI}$\cite{subotnik2013can,dou2017generalized}:
\begin{equation}
    p^{\text{new}} = p - \xi \d_{JI}/|\d_{JI}|,
\end{equation}
where the $\xi$ can be solved by energy conservation:
\begin{equation}\label{kappa}
    \frac{(p_k^{\text{new}})^2}{2}+E_I = \frac{(p_k)^2}{2}+E_J.
\end{equation}
Among the two roots satisfying Eq. \ref{kappa}, we choose the one with a smaller $|\kappa|$.
We now proceed to FSH method in reciprocal space,
and we know that after this canonical transformation from real to reciprocal space in Eqs. \ref{rk} and \ref{pk}, the classical position and momentum become scrambled. Therefore, we can deduce that:
\begin{equation}
\begin{split}
    \bra{\Psi_I^{\text{F(ad)}}}\frac{\partial}{\partial t}\ket{\Psi_J^{\text{F(ad)}}} = &\dot{r}_k\bra{\Psi_I^{\text{F(ad)}}}\frac{\partial}{\partial r_k}\ket{\Psi_J^{\text{F(ad)}}} + \\& \dot{p}_k\bra{\Psi_I^{\text{F(ad)}}}\frac{\partial}{\partial p_k}\ket{\Psi_J^{\text{F(ad)}}}.
\end{split}
\end{equation}
Here, we can define:
\begin{equation}
   d_{IJ}^{r_k} = \bra{\Psi_I^{\text{F(ad)}}}\frac{\partial}{\partial r_k}\ket{\Psi_J^{\text{F(ad)}}},
   d_{IJ}^{p_k} = \bra{\Psi_I^{\text{F(ad)}}}\frac{\partial}{\partial p_k}\ket{\Psi_J^{\text{F(ad)}}}.
\end{equation}
In this way, $\mathbf{P}\cdot\d_{IJ}$ in real-space hopping rate (see Eq. \ref{Gamma_real}) becomes $(p_k\cdot d_{IJ}^{r_k} - \omega^2 r_k\cdot d_{IJ}^{p_k})$.
After a hop from Floquet state $J$ to Floquet state $I$ in reciprocal space, one should rescale not only the nuclear momentum but also the position:
\begin{equation}
\begin{split}
    &p^{\text{new}}_k = p_k - \xi \d_{JI}^{r_k}/|\d_{JI}^{r_k}|, \\ &
    r^{\text{new}}_k = r_k + \xi \d_{JI}^{p_k}/|\d_{JI}^{p_k}|,
\end{split}
\end{equation}
where $\xi$ can be solved by energy conservation:
\begin{equation}
    \frac{(p^{\text{new}}_k)^2}{2}+\frac{\omega^2(r^{\text{new}}_k)^2}{2} +E_I = \frac{(p_k)^2}{2}+\frac{\omega^2(r_k)^2}{2} + E_J.
\end{equation}

\section{Results and discussions}
\subsection{Electronic populations}
\label{sec:population}
In this analysis, we consider a lattice composed of $N = 10$ sites while adhering to the parameters $J_0 = 0.3$, $\hbar\omega = 0.3$, $g = 0.5$, and $T = 0.5$ in the context of the Holstein model and the Peierls model. To obtain these numerical results, we employ a 4th-order Runge-Kutta algorithm to propagate both classical and quantum coordinates. The chosen time step for the algorithm is $dt = 0.01$. 
In this paper, parameters are unitless. When considering room temperature ($0.5 T = 293.15$ $K$) as a reference, the energy unit corresponds to $50.5$ $meV$, the time unit corresponds to $82.7$ $fs$. In addition, we have set the distance between the nearest cites is 5 \AA.
For the sake of statistical accuracy, the outcomes have been averaged across $10,000$ initial thermal conditions that conform to the Boltzmann distribution for the classical coordinates:
\begin{equation}
    P({r_n, p_n}) \propto \prod_n \exp \left( -\frac{1}{2T} (p_n^2 + \omega^2 r_n^2) \right).
\end{equation}
Besides, we set the initial probability that the electron appears in each electronic state to be exactly equal in real space; thus, the probability is completely gathered in the $k=0$ state in reciprocal space.
Here we graph $P_0(t) \equiv \left| \langle 0 | \Psi(t) \rangle \right|^2$ as a function of time.
Without light-matter interactions, for mix quantum-classical dynamics in solid, the original Tully’s SH works reasonably well. This finding has been proven in literature. For instance, in Ref. [\!\citenum{krotz2022reciprocal}], their numerical results show that even without decoherence correction, SH matches well with numerically exact HEOM. With light-matter interactions, the Floquet theory is exact for periodic driving. Such that FSH should work well for laser-dressed solid in general that can be considered as benchmark. 
\begin{figure*}[htbp]
    \centering
    \includegraphics[scale=0.7]{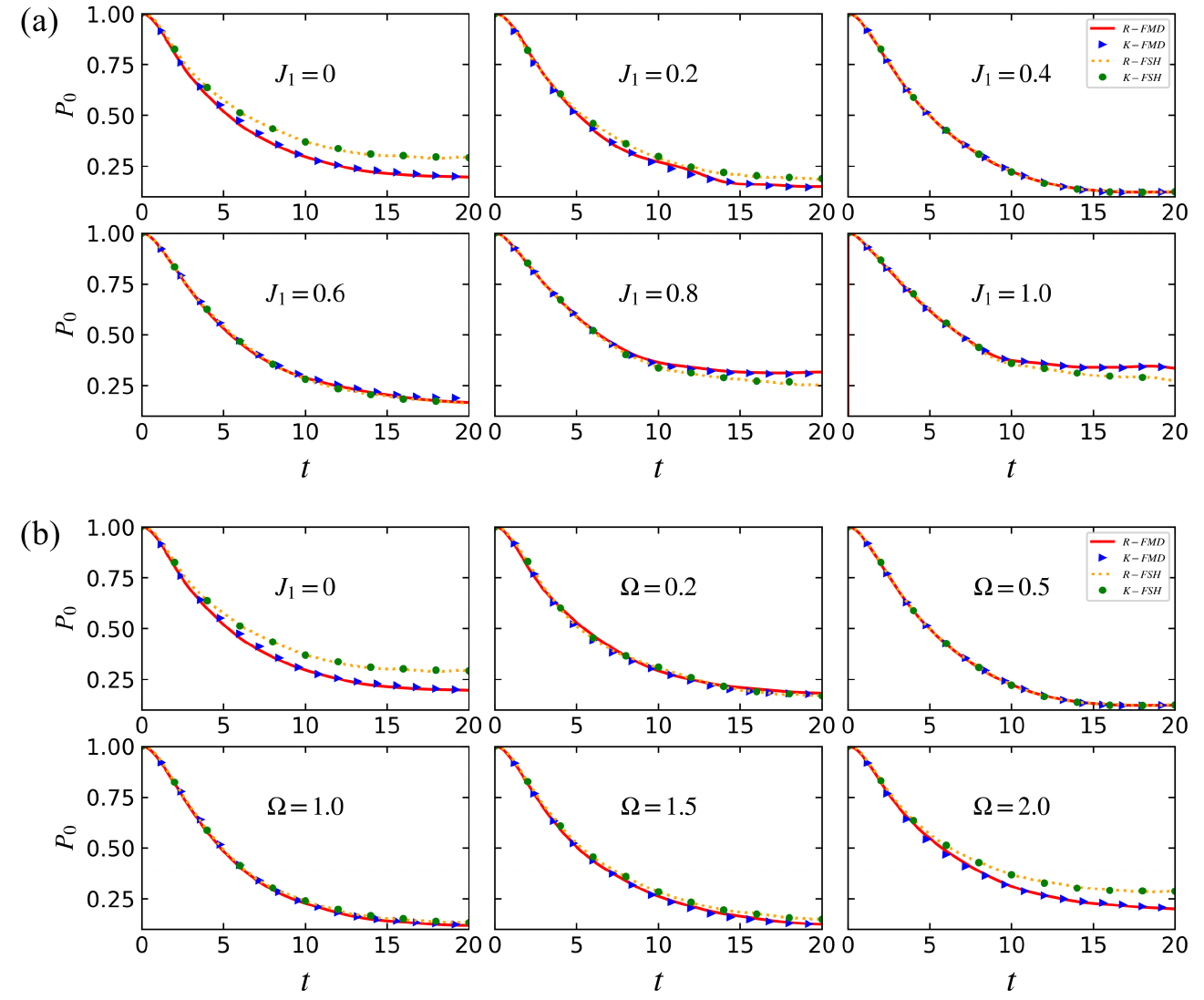}
    \caption{Electronic population dynamics at $k=0$ under varying periodic drivings for the Holstein model, employing four methods: FMD and FSH in both real- and reciprocal-space. In scenario (a), we fix the driving frequency $\Omega=0.5$ and vary the driving amplitude $J_1$ from $0$ to $1$.  In scenario (b), with a constant driving amplitude $J_1=0.4$, the driving frequency $\Omega$ is altered from $0.2$ to $2$. The parameters are set as $J_0=0.3$, $\omega=0.3$, $g=0.5$, $T=0.5$, and $N = 10$. Here, the time unit corresponds to $82.7$ $fs$.}
    \label{fig:pop_Hols}
\end{figure*}

\begin{figure*}[htbp]
    \centering
    \includegraphics[scale=0.7]{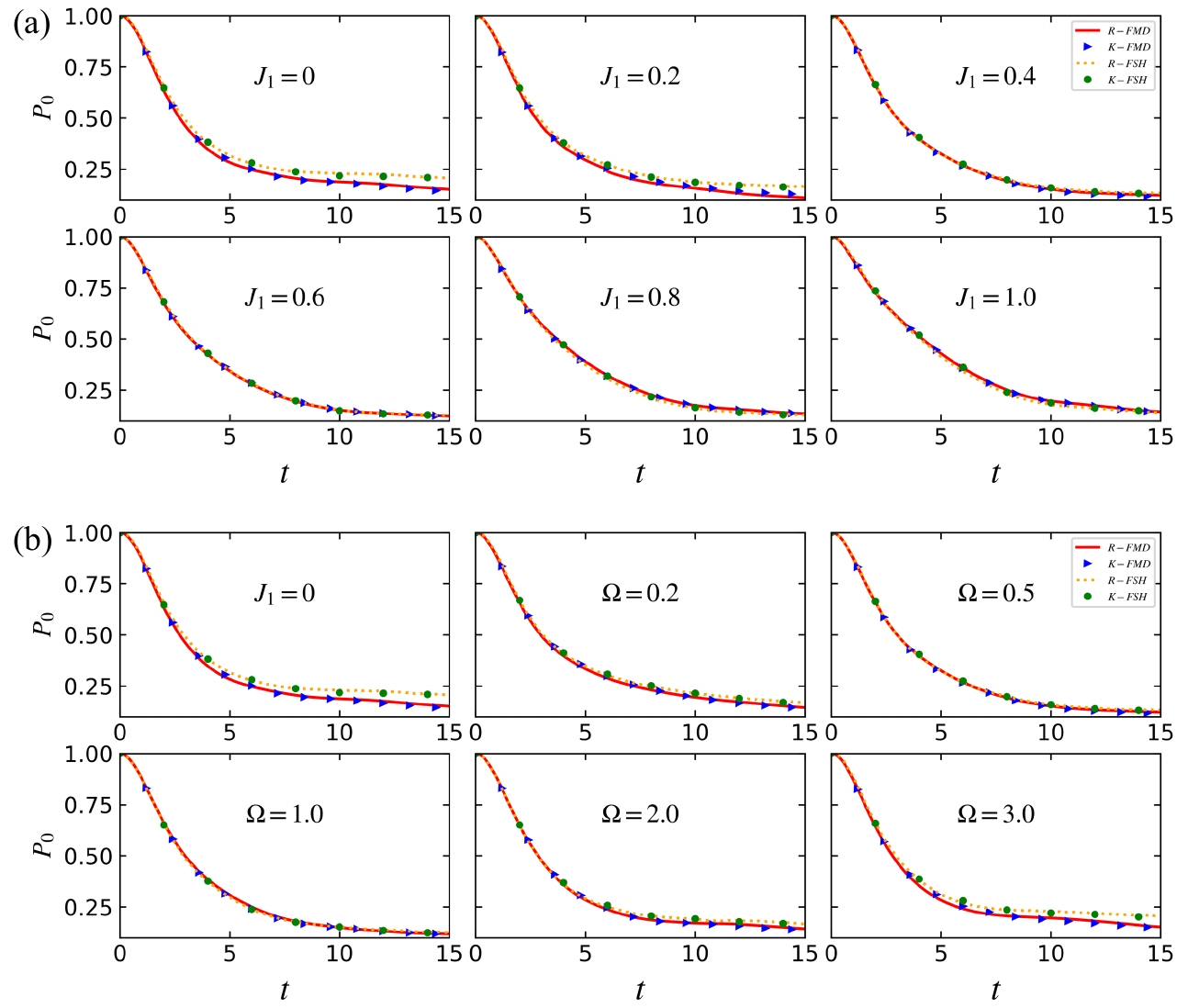}
    \caption{Electronic population dynamics at $k=0$ under varying periodic drivings for the Peierls model, employing four methods: FMD and FSH in both real- and reciprocal-space. In scenario (a), we fix the driving frequency $\Omega=0.5$ and vary the driving amplitude $J_1$ from $0$ to $1$.  In scenario (b), with a constant driving amplitude $J_1=0.4$, the driving frequency $\Omega$ is altered from $0.2$ to $2$. The parameters are set as $J_0=0.3$, $\omega=0.3$, $g=0.5$, and $T=0.5$, and $N = 10$. Here, the time unit corresponds to $82.7$ $fs$.}
    \label{fig:pop_Pei}
\end{figure*}

In Fig. \ref{fig:pop_Hols} (a), we fix the driving frequency $\Omega=0.5$. We observe that with an increase in the driving amplitude $A$, the steady-states of the electronic population at momentum zero initially decrease, followed by a subsequent increase. This observation suggests that periodic drivings have the capacity to transfer energy to electrons within solid-state materials, thereby rendering the electrons more dynamically active.
In Fig. \ref{fig:pop_Hols} (b), with a fixed driving amplitude $A=0.4$, we observe an intriguing pattern as the driving frequency ($\Omega$) is varied. Initially, the steady-states of $\rho_0$ decrease, followed by a subsequent increase until reaching a point where they converge to the same outcome as in the absence of drivings ($J_1=0$). Under fast drivings, where $\Omega$ is sufficiently large, the system may fail to respond to such rapid drivings. Consequently, the properties of solids revert to their original conditions under fast drivings as in the absence of periodic drivings.
Next, we turn to the Peierls model to simulate electronic population dynamics under various periodic drivings, as shown in Fig. \ref{fig:pop_Pei}. Remarkably, we observe a consistent trend reminiscent of the findings in the Holstein model.
For both Holstein and Peierls models, in the absence of light ($J_1=0$), meanfield dynamics (MD) can exhibit incorrect long-time population due to lack of detailed balance. That being said, it is well-known that MD can capture short-time dynamics correctly. On the contrary, SH can predict both short-time and long-time dynamics correctly. With the presence of light-matter interactions, we find that FMD can capture the long-time population well enough in certain regime. 
Still, one can use improved version of the MD method to deal with dynamics in solid state \cite{zheng2023multiple}.

\subsection{Charge mobility simulations}
\begin{figure*}[htbp]
    \centering
    \includegraphics[scale=0.5]{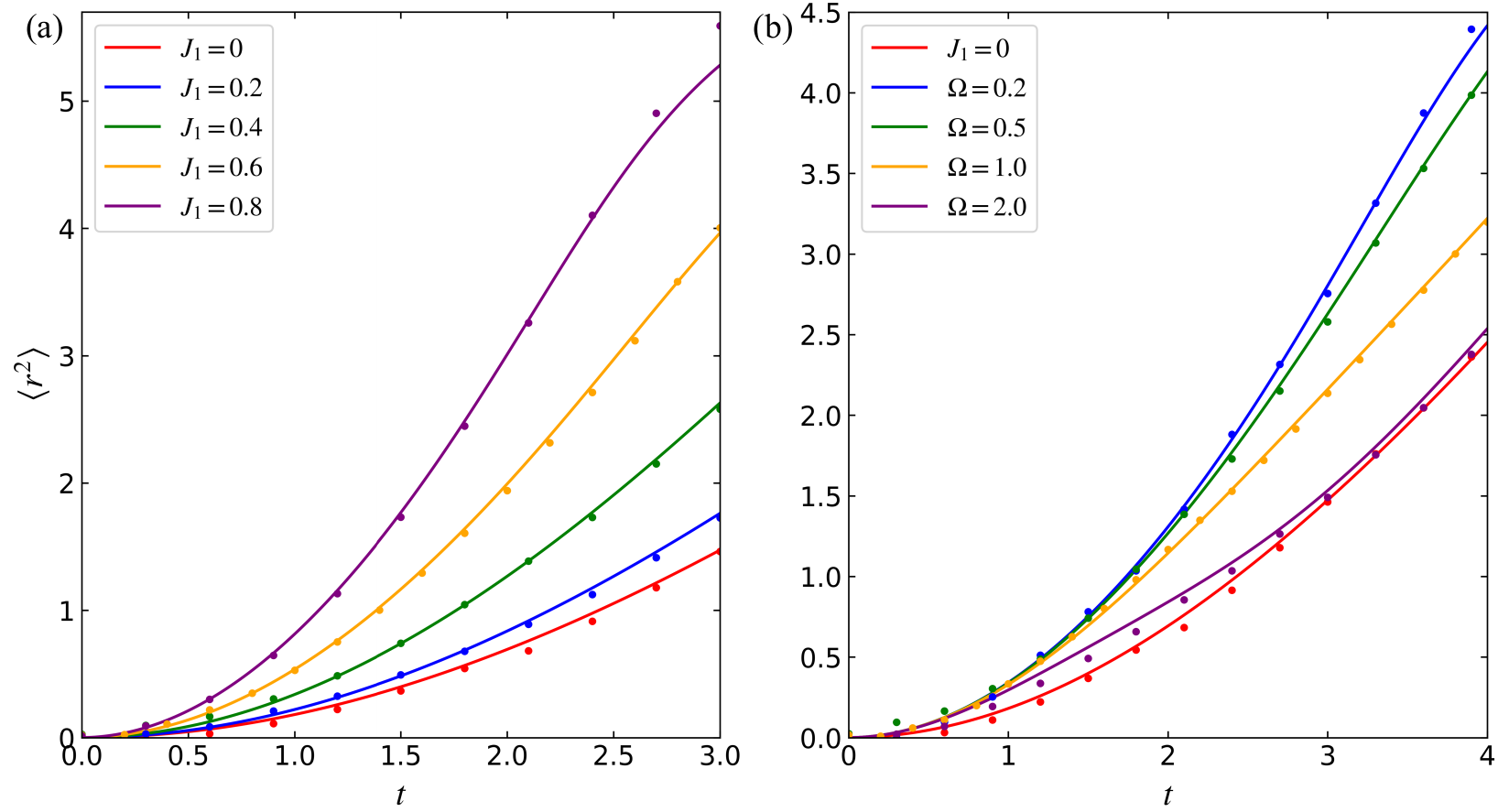}
    \caption{Mean square displacement in real space ($\langle r^2\rangle$) as a function of time under different periodic drivings for Holstein model. a) We fix the driving frequency $\Omega=0.5$, and change the driving amplitude $J_1$ from $0$ to $0.8$. We see $\langle r^2\rangle$ increases with enhancing the driving amplitude $A$. b) We fix the driving amplitude $J_1=0.4$, and change driving frequencies $\Omega$ from $0.2$ to $2$. We see $\langle r^2\rangle$ increases first and then decreases to a point where there is no driving. Here, $J_0=0.3$, $\omega=0.3$, $g^2=0.25$, $T=0.5$, and $N = 10$. The distance unit corresponds to $5$ \AA, and the time unit corresponds to $82.7$ $fs$.}
    \label{fig:mobility_Hols}
\end{figure*}

\begin{figure*}[htbp]
    \centering
    \includegraphics[scale=0.5]{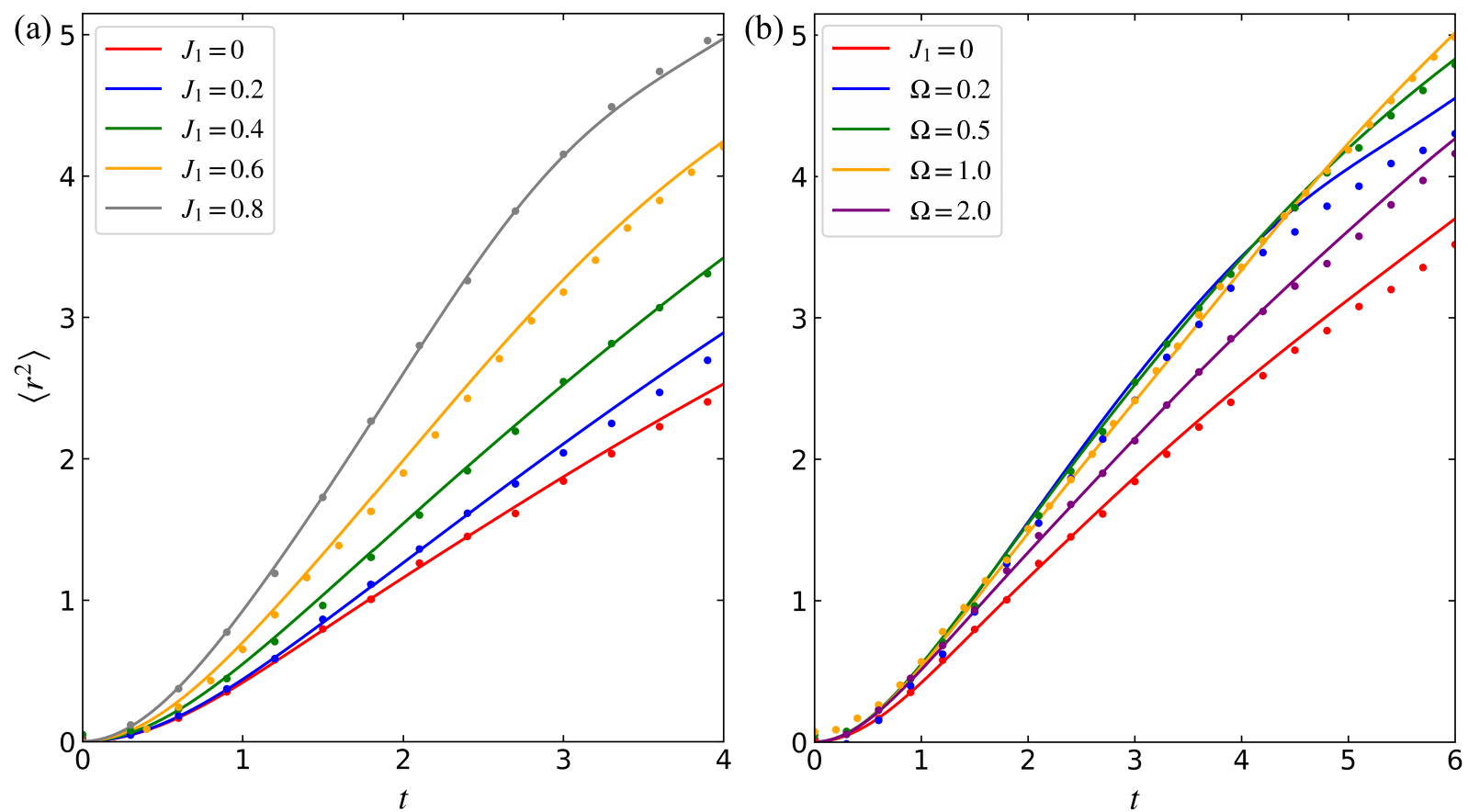}
    \caption{Mean square displacement in real space ($\langle r^2\rangle$) as a function of time under different periodic drivings for Holstein model. a) We fix the driving frequency $\Omega=0.5$, and change the driving amplitude $J_1$ from $0$ to $0.8$. We see $\langle r^2\rangle$ increases with enhancing the driving amplitude $A$. b) We fix the driving amplitude $J_1=0.4$, and change the driving frequencies $\Omega$ from $0.2$ to $2$. We see $\langle r^2\rangle$ increases first and then decreases to a point where there is no driving. Here, $J_0=0.3$, $\omega=0.3$, $g^2=0.25$, $T=0.5$, and $N = 10$. The distance unit corresponds to $5$ \AA, and the time unit corresponds to $82.7$ $fs$.}
    \label{fig:mobility_Peil}
\end{figure*}

By employing our FMD and FSH methods, we also simulate charge mobility in a short time under different periodic drivings in real space, as shown in Figs. \ref{fig:mobility_Hols} and \ref{fig:mobility_Peil}.
Short-time charge mobility is essential for semiconductor devices or high-frequency electronics. For example, gallium nitride (GaN) high electron-mobility transistors have emerged as excellent power devices\cite{meneghesso2008reliability,meneghini2021gan}. 
Here, we simulate the dynamics of the mean square displacement $\langle r^2\rangle$ of the charge in real space, which is proportional to the charge mobility\cite{wang2010computational}.
Different from \ref{sec:population}, here we initialize the electronic population at the middle site in real space and take an average of $10,000$ trajectories for each dynamic result.

For the Holstein model, we find that by adding Floquet drivings, short-time charge mobility can be enhanced with increasing the strength of the driving (see Fig. \ref{fig:mobility_Hols} (a)). When the driving frequency becomes sufficiently high, the system may fail to respond effectively to the rapid driving, resulting in charge mobility comparable to the condition without periodic drivings.
In the Peierls model, we observe a similar trend to that in the Holstein model, as shown in Fig. \ref{fig:mobility_Peil}.
These numerical results illustrate two key points: firstly, we can induce charge migration in solid materials by progressively increasing the light intensity; secondly, when the light intensity is fixed, it is essential to identify the optimal intensity (neither too high nor too low) to maximize charge mobility.

\subsection{Truncation in Brillouin zone}
The advantage of using a real-space approach lies in its ability to precisely capture local interactions within a specific spatial region while keeping computational demands in check by limiting the scope of local interactions considered. Similarly, for phenomena restricted to a certain range within the Brillouin zone, a comparable strategy can be employed in reciprocal space. In scenarios where electronic carriers are localized due to either the short time scales not permitting full Brillouin zone traversal or their stabilization at band minima, such confinement can be effectively modeled. This concept has been effectively employed in the theoretical analysis of exciton and trion states in monolayer transition-metal dichalcogenides.\cite{PhysRevB.93.235435} In these studies, the analysis was focused around the K points in the Brillouin zone by implementing a specific truncation radius, thereby streamlining the computations.

In FMD and FSH dynamics, a reciprocal space formulation allows for effective truncation when the initial population is concentrated at or near $k = 0$. Here we set $N = 20$, $J_0 = 0.6$, $\hbar\omega = 0.2$, $J_1 = 0.3$, $\Omega = 0.5$, $g^2 = 5$, and $T = 1$. 
We symmetrically truncate the corner of the Brillouin zone and keep $4/5$, $3/5$, and $2/5$ sizes of the original one, and the corresponding size numbers $N$ are $16$, $12$, and $8$, respectively.
We emphasize again that truncation in reciprocal space does save a lot of time computational time. Theoretically speaking, the computational time of surface hopping or Ehrenfest dynamics scales as $O(N^3)$ as a function of $N$, where $N$ is the number of the quantum states. In our simulation, we find that we can use as less as 60\% of the total $k$ states in both FSH and FMD methods to recover the results without any truncation. Such that we can save 80\% of the total computational time.

\begin{figure*}[htbp]
    \centering
    \includegraphics[scale=0.65]{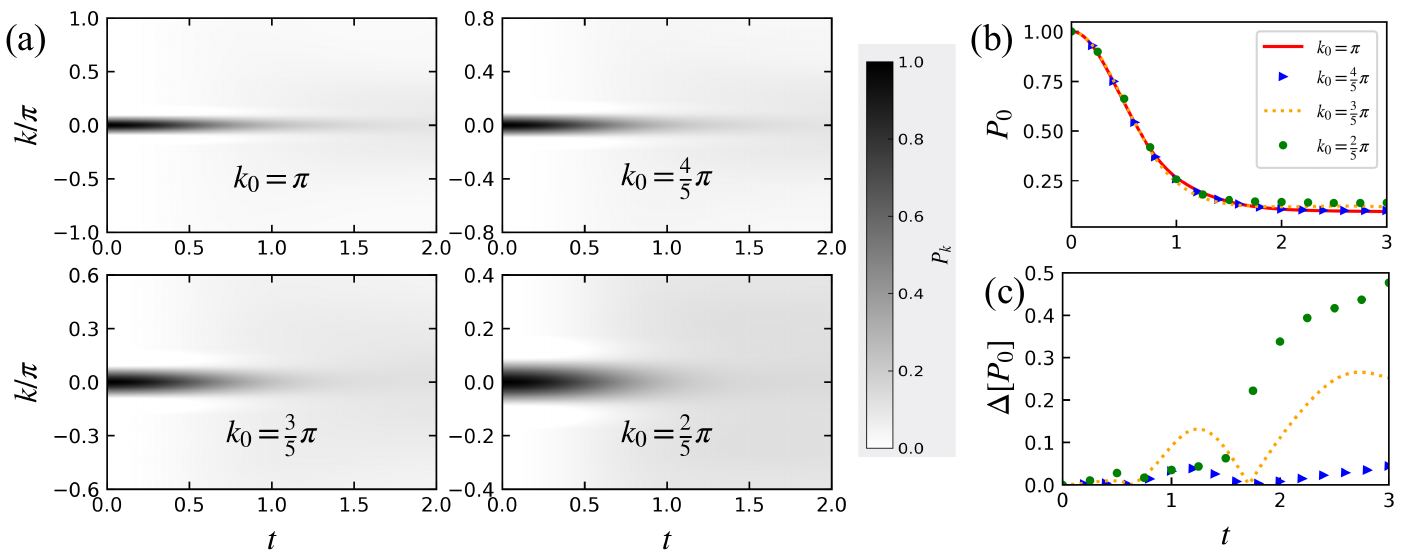}
    \caption{Truncated Electronic Population Dynamics $P_k(t)$ for the Holstein Model Using the FMD Method.
a) Comparison of the truncated overall population dynamics over time against the untruncated results.
b) Time evolution of the truncated $P_0(t)$ at the center of the Brillouin zone, juxtaposed with the untruncated counterpart.
c) Analysis of the error $\Delta P_0(t)$, quantifying the deviation from the full Brillouin zone results, across various truncation radii $k_0$. Parameters: $J=0.6$, $\omega=0.2$, $J_1=0.3$, $\Omega=0.5$, $g^2=5$, $T=1$, and $N = 20$. Here, the time unit corresponds to $82.7$ $fs$.}
    \label{fig:trun_Heis_mf}
\end{figure*}

\begin{figure*}[htbp]
    \centering
    \includegraphics[scale=0.65]{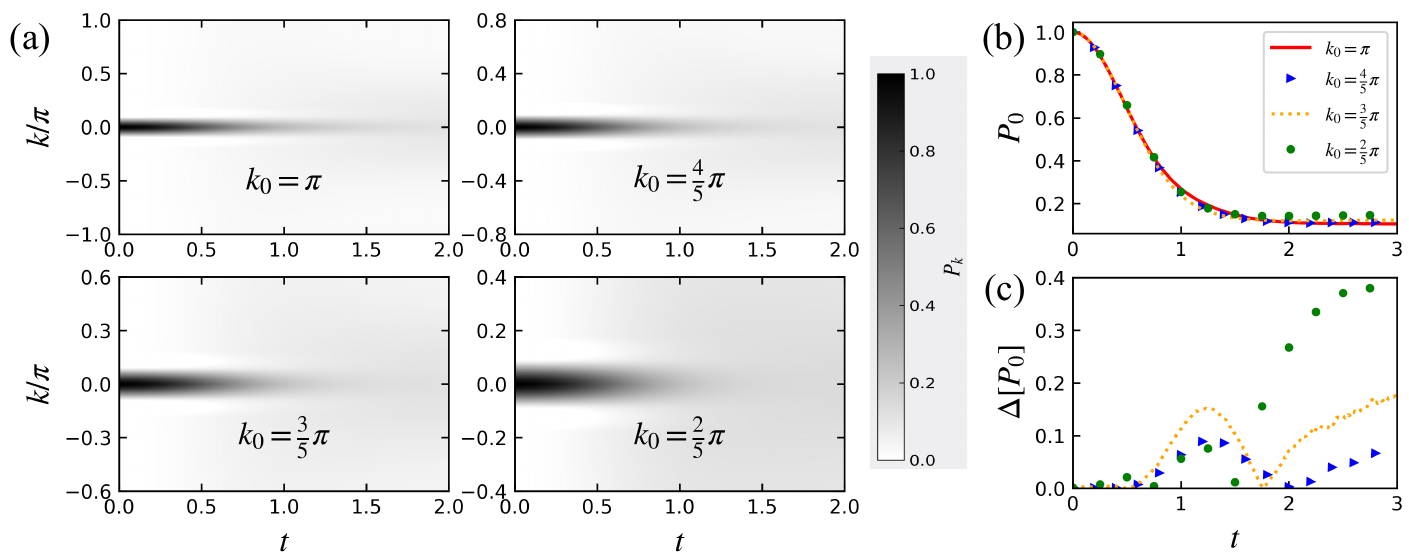}
    \caption{Truncated Electronic Population Dynamics $P_k(t)$ for the Holstein Model Using the FSH Method.
a) Comparison of the truncated overall population dynamics over time against the untruncated results.
b) Time evolution of the truncated $P_0(t)$ at the center of the Brillouin zone, juxtaposed with the untruncated counterpart.
c) Analysis of the error $\Delta P_0(t)$, quantifying the deviation from the full Brillouin zone results, across various truncation radii $k_0$. Parameters: $J_0=0.6$, $\omega=0.2$, $J_1=0.3$, $\Omega=0.5$, $g^2=5$, $T=1$, and $N = 20$. Here, the time unit corresponds to $82.7$ $fs$.}
    \label{fig:trun_Heis_sh}
\end{figure*}

\begin{figure*}[htbp]
    \centering
    \includegraphics[scale=0.65]{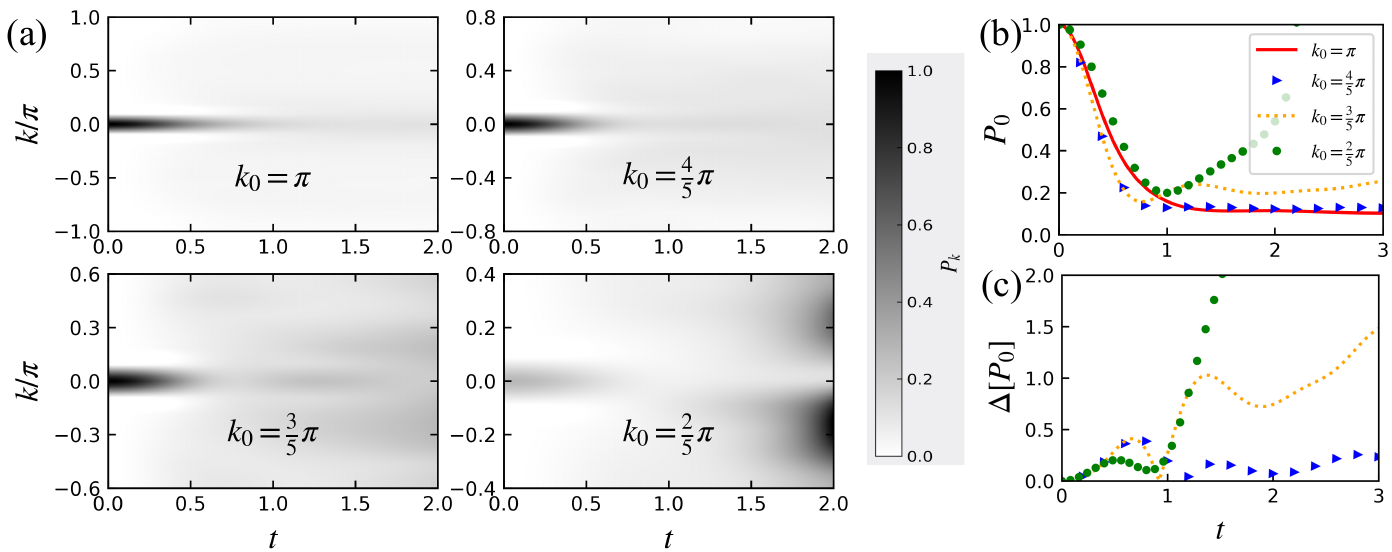}
    \caption{Truncated Electronic Population Dynamics $P_k(t)$ for the Peierls Model Using the FMD Method.
a) Comparison of the truncated overall population dynamics over time against the untruncated results.
b) Time evolution of the truncated $P_0(t)$ at the center of the Brillouin zone, juxtaposed with the untruncated counterpart.
c) Analysis of the error $\Delta P_0(t)$, quantifying the deviation from the full Brillouin zone results, across various truncation radii $k_0$. Parameters: $J=0.6$, $\omega=0.2$, $J_1=0.3$, $\Omega=0.5$, $g^2=5$, $T=1$, and $N = 20$. Here, the time unit corresponds to $82.7$ $fs$.}
    \label{fig:trun_Perl_mf}
\end{figure*}

\begin{figure*}[htbp]
    \centering
    \includegraphics[scale=0.65]{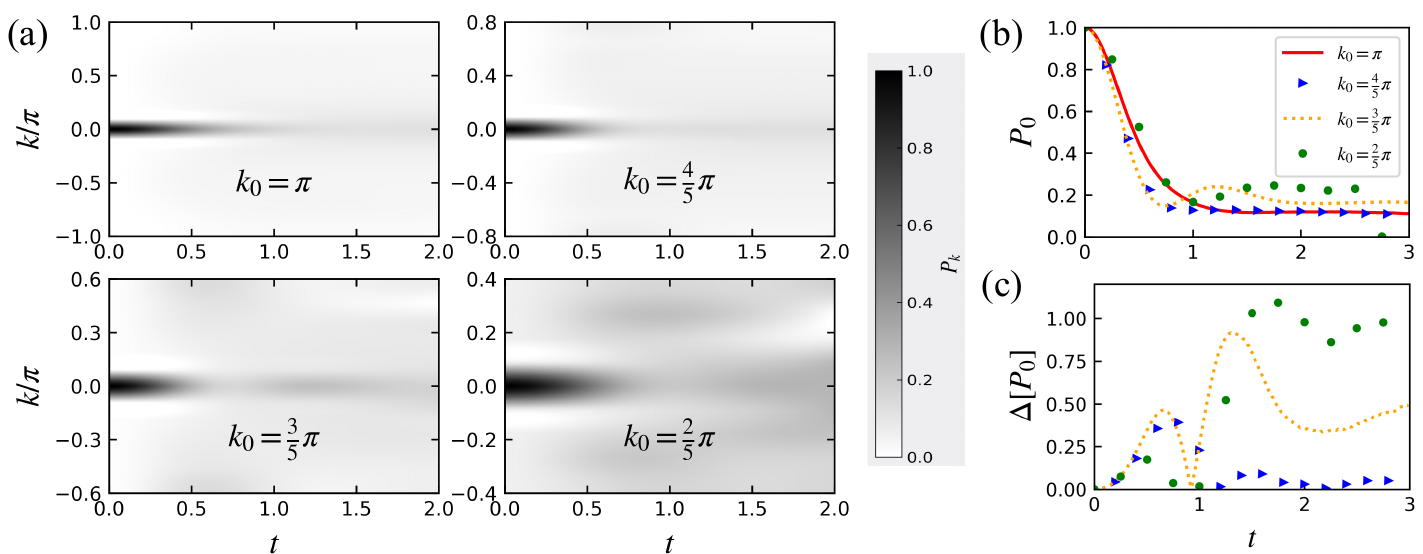}
    \caption{Truncated Electronic Population Dynamics $P_k(t)$ for the Peierls Model Using the FSH Method.
a) Comparison of the truncated overall population dynamics over time against the untruncated results.
b) Time evolution of the truncated $P_0(t)$ at the center of the Brillouin zone, juxtaposed with the untruncated counterpart.
c) Analysis of the error $\Delta P_0(t)$, quantifying the deviation from the full Brillouin zone results, across various truncation radii $k_0$. Parameters: $J=0.6$, $\omega=0.2$, $J_1=0.3$, $\Omega=0.5$, $g^2=5$, $T=1$, and $N = 20$. Here, the time unit corresponds to $82.7$ $fs$.}
    \label{fig:trun_Perl_sh}
\end{figure*}

Figs. \ref{fig:trun_Heis_mf} and \ref{fig:trun_Heis_sh} illustrate the temporal evolution of the whole population (Figs. \ref{fig:trun_Heis_mf} (a) and \ref{fig:trun_Heis_sh} (a)) and $P_0$ (Figs. \ref{fig:trun_Heis_mf} (b) and \ref{fig:trun_Heis_sh} (b)) for the Holstein model under different truncation scales, as calculated by FMD and FSH respectively. Remarkably, these methods yield accurate electronic population profiles even with significant truncation, such as at $k_0 = \frac{2}{5}\pi$, closely matching the untruncated results, as can be seen in the Figs. \ref{fig:trun_Heis_mf} (c) and \ref{fig:trun_Heis_sh} (c).
However, as observed in Fig. \ref{fig:trun_Perl_mf}, the accuracy of large truncations ($k_0 = \frac{2}{5}\pi$) in the FMD method for the Peierls model is less satisfactory. In contrast, as shown in Fig. \ref{fig:trun_Perl_sh}, the FSH method demonstrates superior performance under similar truncation conditions, highlighting its advantages.
Importantly, truncation calculations is particularly beneficial when dealing with large systems, especially for the Floquet Hamiltonian, making it an attractive strategy for efficiently exploring electronic dynamics in complex models.

\section{Conclusions}
Our study successfully constructs Floquet Hamiltonian for the Holstein and Peierls models, showing the huge impact of Floquet driving on Floquet replicas.
The derived FMD and FSH, explored in both real and reciprocal space, yield consistent results through numerical simulations. This agreement validates the reliability of our theoretical framework, emphasizing the equivalence of real and reciprocal space outcomes.
The advantage of simulations in real space is that the algorithm is simple and easy to implement. In addition, simulations in real space are generally more suitable for disordered solids, where the periodicity in real space is broken such that the Block theory is invalid. On the contrary, simulations in reciprocal space provides more physical insights on exciton dynamics in band structure. The band structure can be obtained from electronic structure theory, such that the dynamics in reciprocal space can combine with electronic structure calculation directly. Furthermore, the truncation can be done for simulations in reciprocal space, which saves a lot of computational time.
However, FSH is inefficient when dealing with slow drivings due to the large Floquet replicas. The applications of FMD are limited but efficient for large system.

Significantly, our results showcase that light-matter interactions, facilitated by Floquet driving, can markedly enhance electronic mobility. This finding points towards practical applications, suggesting the possibility of engineering materials with improved conductivity. Moreover, our study demonstrates the potential for computational efficiency through truncation techniques, which are particularly effective when applied to well-chosen initial conditions. This insight opens avenues for reducing computational costs in simulating Floquet-driven systems.

In conclusion, our research contributes valuable insights into the Floquet-driven dynamics of fundamental models, bridging theoretical understanding with practical applications. These findings hold promise for the future design and optimization of materials with tailored electronic properties.

\begin{acknowledgments}
This material is based upon work supported by National Natural Science Foundation of China (NSFC No. 22273075). W.D. acknowledges start-up funding from Westlake University. 
\end{acknowledgments}

\bibliography{reference}
\end{document}